# Angular momentum and the polar basis of harmonic oscillator

**M. Hage-Hassan**

Université Libanaise, Faculté des Sciences Section (1)
Hadath-Beyrouth

## Abstract

   In this paper we follow the Schwinger approach for angular momentum but with the polar basis of harmonic oscillator as a starting point. We derive by a new method two analytic expressions of the elements of passage matrix from the double polar basis to 4-dimensions polar basis of the harmonic oscillator. These expressions are functions of the modules of magnetic moments. The connection between our results and the results derived by the group theory of Laguerre polynomials is found. We determine a new expression for these elements in terms of magnetic moments in the general case. We deduce from these expressions the symmetries of 3j symbols. A new generating function of the Clebsh-Gordan coefficients, functions of the modules of magnetic moments are found. We prove that the generating function of recoupling coefficients 3nj for the polar basis are the same in the Schwinger's approach therefore the polar basis of harmonic oscillator may be a starting point to study the angular momentum.

## 1-Introduction

   The theory of angular momentum occupies an important position in the development of all physical theories of atomic physics, nuclear physics and particle physics. We observe in this theory that the Wigner's D-matrix elements are functions of Jacobi polynomials $P_n^{(\alpha,\beta)}$ with positive indices, α and β, but all the results derived from it are valid without this restriction by making use of the symmetry properties [1]. The extension of this idea to Clebsh-Gordan coefficients remain to be proven in the general case.

   We find also in the famous Schwinger's work "on angular momentum" [2] a search of a generating function for 3j symbols function of modules of magnetic moments but the function obtained was with three parameters are not usable for obtaining a simple expression of these symbols. On the other hand in many bases [3-6] there are also other useful physical polynomials with restrictive indices and the matrix elements of passage between the bases has symmetry and may be calculated without restriction. So the problem is more general and occurs in many problems of quantum physics.

   To verify this point of view we choose the polar basis of harmonic oscillator (2dPH) as a basis, unlike Schwinger abstract basis, for the study of angular momentum [6, 8]. The double polar basis $|(4dPH)\rangle$ is function of the product of Jacobi polynomials for the angular part and Laguerre polynomials for the radial; this allows us to calculate the matrix elements $\langle 2(2dPH)|(4dPH)\rangle$ using the orthogonality of these polynomials. We propose two different methods and we derive two expressions for the restrictive indices.

We perform also the calculation for unrestricted indices with a new generating function for the polar basis (2dPH) so we get new expressions of these coefficients. The identification of these expressions shows that the restriction of these indices is due to the symmetry. We find the integral representation of Clebsh-Gordan coefficients in terms of Laguerre polynomial then we deduce a new generating function dependant of two parameters and from which we can deduce the analytical expression. We prove the Relation between our results and the results derived by the group theory of Laguerre polynomials [7].We prove also that the recoupling coefficients are the same of Schwinger's approach therefore they are independent of the choice of the polar basis. Applying our method to classical groups will be the subject of another paper.

The plan of the paper is as follows. We begin, in section 2, by reviewing the basis of (2dPH) and 4-dimension polar basis of the harmonic oscillator. In section 3, we give a derivation of the generating function for the passage matrix elements between bases in terms of the module of magnetic moments. We expose in section 4 the relation of our method and the group theory of Laguerre polynomials. In section 5 we find the Generating functions of passage elements from 2(2dPH) to 4-dimensions polar basis in terms of magnetic moments (general case). Section 6 is devoted to resume our results on the symmetry of 3j symbols and the new generating functions of Clebsh-Gordan coefficients. In Section 6 we prove the connection between Schwinger approach for 6j, 9j symbols and the polar basis of harmonic oscillator.

## 2. The basis of (2dPH) and 4-dimension polar basis of harmonic oscillator

Using the method of separation of variables we can find the polar basis and the 4-dimensional polar basis of harmonic oscillator. We shall give only the expressions of these bases [9].

### 2.1- The basis of 2(2dPH) $(\rho_1, \rho_2, \alpha, \beta)$

To determine the eigenfunctions of two dimensions harmonic oscillator we put

$$u_1 = \rho_1 \cos\alpha, \quad u_3 = \rho_2 \cos\beta$$
$$u_2 = \rho_1 \sin\alpha, \quad u_4 = \rho_2 \cos\beta \qquad (2.1)$$
$$0 \le \rho_1 \le \infty, 0 \le \rho_2 \le \infty, 0 \le \alpha \le 2\pi, 0 \le \beta \le 2\pi,$$

The solution of Schrödinger equation is:

$$\Phi_{jm}(\rho,\alpha) = f_{jm}(\rho)e^{-2im\alpha} =$$
$$\frac{1}{\sqrt{\pi}}\sqrt{\frac{(j-|m|)!}{(j+|m|)!}}\exp(-\frac{\rho^2}{2})L_{j-|m|}^{2|m|}(\rho^2)\rho^{2|m|}e^{-2im\alpha} \qquad (2.2)$$

With $\qquad E_{nm} = (2n+2|m|+1)\hbar\omega$ and $j = n+|m|$ $\qquad (2.3)$

$$\int_0^\infty e^{-x}x^\alpha (L_n^\alpha(x))^2 dx = \frac{\Gamma(\alpha+n+1)}{n!} \qquad (2.4)$$

The double polar basic of harmonic oscillator 2(2dPH) is

$$\Phi_{j_1 m_1}(\rho_1, \alpha_1)\Phi_{j_2 m_2}(\rho_2, \alpha_2) = f_{j_1 m_1}(\rho_1) f_{j_2 m_2}(\rho_2) e^{-2im_1\alpha_1} e^{-2im_2\alpha_2} \quad (2.5)$$

With $\quad j_1 = n_1 + |m_1|, \quad j_2 = n_2 + |m_2| \quad (2.6)$

### 2.2- The 4-dimensions polar basis (4dPH) $(u, \psi, \theta, \varphi)$

We choose a new parameterization of the harmonic oscillator basis (4dPH)

Put $\quad \rho_1 = r\cos(\theta/2), \rho_2 = r\sin(\theta/2) \quad (2.7)$

We write $\quad u_1 = \rho_1 \cos(\alpha), u_2 = \rho_1 \sin(\alpha),$

$$u_3 = r\rho_2 \cos(\beta), u_4 = \rho_2 \sin(\beta) \quad (2.8)$$

With $\quad \alpha = (\varphi - \psi)/2, \beta = (\psi + \varphi)/2$, $\psi$, $\theta$ and $\varphi$ are Euler's angles.

$$0 \leq \theta \leq \pi/2, \; 0 \leq \psi \leq 2\pi, \; 0 \leq \varphi \leq 2\pi, \; d\vec{r} = r^3 dr d\Omega / 8\pi^2$$

With these notations we can go to the dimensions 8, 16, 32... [21].
The eigenfunctions of the harmonic oscillator (4dPH) [1-3, 7] is:

$$\Psi_{(n)j_3 m_3 mm'} = (1/\pi) N_{n,j_3} r^{(2j_3)} D^{j_3}_{(m,m')}(\Omega) e^{-r^2/2} L_n^{2j_3+1}(r^2). \quad (2.9)$$

The conservation of energy implies that:

$$2j_3 + 2n = 2j_1 + 2j_2$$

We find $\quad n = j_1 + j_2 - j_3, \; \alpha = 2j_3 + 1$

$$N_{n,j_3} = \left[\frac{(j_1 + j_2 - j_3)!}{(j_1 + j_2 + j_3 + 1)!}\right]^{\frac{1}{2}} \quad (2.10)$$

And $\quad D^{j_3}_{(m,m')}(\Omega) = e^{-i2|m_2|\alpha - i2|m_1|\beta} d^{j_3}_{(m,m')}(\theta),$

With $\quad d^{j_3}_{(m,m')}(\theta) = \left[\frac{(2j_3+1)(j_3+m)!(j_3-m)!}{(j_3+m')!(j_3-m')!}\right]^{1/2} \times$

$$(\cos\frac{\theta}{2})^{2|m_2|}(\sin\frac{\theta}{2})^{2|m_1|} P^{(2|m_2|,2|m_1|)}_{j_3-m}(\cos\theta) \quad (2.11)$$

And $\quad 2|m_1| = m - m', \; 2|m_2| = m + m', \; m = |m_1| + |m_2|, \; m' = |m_2| - |m_1|$

We deduce that

$$\Psi_{(n)j_3 m'm}(r,\psi\theta\varphi) = (1/\pi)\left[\frac{(2j_3+1)(j_1+j_2-j_3)!(j_3+m)!(j_3-m)!}{(j_1+j_2+j_3+1)!(j_3+m')!(j_3-m')!}\right]^{1/2} \times$$

$$\rho_1^{2|m_1|} \rho_2^{2|m_2|} r^{2(j_3-|m_1|-|m_2|)} L_n^{2j_3+1}(r^2) e^{-r^2} P^{(2|m_2|,2|m_1|)}_{j_3-m}(\cos(\theta)) e^{-2i(m_1\alpha + m_2\beta)}. \quad (2.12)$$

And $\quad \cos(\theta) = (\rho_1^2 - \rho_2^2)/(\rho_1^2 + \rho_2^2)$

### 3. Expression and generating function of the passage matrix elements between bases in terms of the module of magnetic moments

We observe that the functions of the basis, the radial and the angular are orthogonal polynomials so we can use either one or the other of these functions for calculating the matrix elements of passage $\langle \Phi_{j_1 m_1}, \Phi_{j_2 m_2} | \Psi_{n j_3 m'm} \rangle$.

## 3.1 Elements of the passage matrix (first method)

We can do the calculation by the generating functions methods of Laguerre polynomials [10, 18-20] but we will just do the calculations by a new and much simpler method using the developments of these polynomials.

Starting from the development

$$\Phi_{j_1 m_1}(\rho_1, \alpha_1)\Phi_{j_2 m_2}(\rho_2, \alpha_2) = \sum_{j_3} \langle \Psi_{n j_3 m' m} | \Phi_{j_1 m_1}, \Phi_{j_2 m_2} \rangle \Psi_{n j_3 m' m}(r, \psi\theta\varphi) \quad (3.1)$$

We deduce that the elements of the matrix are:

$$\langle \Psi_{n j_3 m' m} | \Phi_{j_1 m_1} \Phi_{j_2 m_2} \rangle = \int d\vec{r} \Phi_{j_1 m_1}(\rho_1, \alpha_1)\Phi_{j_2 m_2}(\rho_1, \alpha_2) \Psi_{n j_3 m' m}(r, \psi\theta\varphi) \quad (3.2)$$

### 3.1.1 Integration of radial part

We know that any polynomial of degree n can be written in ascending order from zero to n or descending order from n to zero. We will use the second form in the development of Laguerre polynomials [13]:

$$L_n^\alpha(x) = (-1)^n \sum_{k=0}^n (-1)^k \frac{\Gamma(\alpha+n+1)}{k!(n-k)!\Gamma(\alpha+n-k+1)} x^{n-k} \quad (3.3)$$

We write

$$L_{n_1}^{2|m_1|}(r^2 \cos^2\theta/2) L_{n_2}^{2|m_2|}(r^2 \sin^2\theta/2) =$$

$$\sum_{i=0}^{n_1}\sum_{j=0}^{n_2}(-1)^{i+j}[\Delta_{n_1,n_2,i,j}^{2|m_1|,2|m_2|}\cos^{2(n_1-i)}(\theta/2)\sin^{2(n_2-j)}(\theta/2)]r^{2(n_1-i)+2(n_2-j)} \quad (3.4)$$

The power of r is

$$2(n_1-i)+(2n_2-j)+2|m_1|+2|m_2| = 2(j_1+j_2-j_3-i-j)+2j_3+1$$

Using the formula [14]

$$\frac{x^r}{r!} = \sum_{i=0}^r (-1)^i \binom{r+\alpha}{r-i} L_i^{(\alpha)}(x) \quad (3.5)$$

We find

$$\frac{(r^2)^{j_1+j_2-i-j-j_3}}{(j_1+j_2-i-j-j_3)!} = \sum_{k=0}^n (-1)^k \binom{j_1+j_2-i-j-j_3+j_3+1}{j_1+j_2-i-j-j_3-k} L_k^{(2j_3+1)}(r^2) \quad (3.6)$$

With

$$j_1+j_2-i-j-j_3-k \geq 0.$$

Orthogonality of the generalized Laguerre polynomials (2.4) implies that k must be equal to n ($n = j_1 + j_2 - j_3$) for the integral (3.2) be different from zero. We deduce that

$$j_1+j_2-i-j-j_3-k = -i-j \geq 0$$

But i and j are positive numbers we thus infer that i = j = 0. We then do the integration on the radial part with (2.4).

### 3.1.2 Integration of angular part and matrix elements

We evaluate the integral of angular part of (3.2) with the help of integral over the product of three D's or the Gaunt integral [1, 15] in terms of Clebsh-Gordan:

$$\langle ab, \alpha\beta | (ab)c\gamma \rangle = (-1)^{b+\beta} \left[ \frac{(\Delta+1)!(\Delta-2c)!}{(a+\alpha)!(a-\alpha)!(b+\beta)!(b-\beta)!} \right]^{\frac{1}{2}} \times$$

$$\int_0^\pi (\cos(\frac{\theta}{2})\sin(\frac{\theta}{2}))^{a+b} (tg(\frac{\theta}{2}))^{\beta-\alpha} d^c_{(\gamma,b-a)}(\theta)\sin\theta\, d\theta, \Delta = a+b+c, \alpha+\beta+\gamma=0$$

We obtain $\left\langle \Phi_{j_1 m_1}, \Phi_{j_2 m_2} \middle| \Psi_{n j_3 m' m} \right\rangle = e^{i\pi\varphi} \left\langle j_1^{|m|} j_2^{|m|}, m_1^{|m|} m_2^{|m|} \middle| (j_1^{|m|} j_2^{|m|}) j_3^{|m|} m_3^{|m|} \right\rangle$, (3.7)

And
$$j_1^{|m|} = (j_1 + j_2 - |m_1| + |m_2|)/2,\; m_1^{|m|} = (j_2 - j_1 + |m_1| + |m_2|)/2,$$
$$j_2^{|m|} = (j_1 + j_2 + |m_1| - |m_2|)/2,\; m_2^{|m|} = (-j_2 + j_1 + |m_1| + |m_2|)/2$$
$$j_3^{|m|} = j_3,\; m_3^{|m|} = (|m_1| + |m_2|)\quad \varphi = (j_2^{|m_2|} + m_2^{|m|} + N), N = n_1 + n_2 + n$$

Analogue result is already obtained by Mardoyan and al. [3] and [4-5] by an approximation method.

### 3.2 Elements of the passage matrix using the radial functions (Second method)

The method of calculation of these elements using the radial functions is very useful for many applications [11-12]. We start from the expression

$$\Psi_{n j_3 m' m}(r, \psi\theta\varphi) = \sum_{j_1 j_2 m_1 m_2} \left\langle \Phi_{j_1 m_1}, \Phi_{j_2 m_2} \middle| \Psi_{n j_3 m' m} \right\rangle \Phi_{j_1 m_1}(\rho_1, \alpha_1) \Phi_{j_2 m_2}(\rho_2, \alpha_2) \quad (3.8)$$

We find that $(1/\pi) \left[ \dfrac{(2j_3+1)(j_1+j_2-j_3)!(j_3+m)!(j_3-m)!}{(j_1+j_2+j_3+1)!(j_3+m')!(j_3-m')!} \right]^{1/2} \times$

$$\rho_1^{2|m_1|} \rho_2^{2|m_2|} r^{2(j_3-|m_1|-|m_2|)} L_n^{2j_3+1}(r^2) e^{-r^2} P_{j_3-m}^{(|k_1|,|k_2|)}(\cos(\theta)) e^{-ik_1\alpha - ik_2\beta} =$$

$$\{\frac{1}{\pi} \sum_{n_1 n_2, m_1, m_2} \left\langle \Phi_{j_1 m_1}, \Phi_{j_2 m_2} \middle| \Psi_{n j_3 m' m} \right\rangle \sqrt{\frac{(j_1-|m_1|)!(j_2-|m_2|)!}{(j_1+|m_1|)!(j_2+|m_2|)!}} \times$$

$$\exp(-\frac{\rho_1^2 + \rho_2^2}{2}) \rho_1^{2|m_1|} \rho_2^{2|m_2|} L_{n_1}^{2|m_1|}(\rho_1^2) L_{n_2}^{2|m_2|}(\rho_2^2) e^{-2im_1\alpha_1} e^{-2im_2\alpha_2} \} \quad (3.9)$$

But $m = (m_1 + m_2), m' = (m_1 - m_2)$ then $m_1, m_2, n_1$ and $n_2$ are fixed. If $\rho_2 \neq 0$ we simplify by $\rho_1^{2|m_1|} \rho_2^{2|m_2|}$ the both sides of the expression (3.3) and then we put $\rho_2 = 0$, $x = \rho_1^2$ and using the expressions [18]

$$P_{(j_3-m)}^{(2|m_1|,2|m_2|)}(1) = \binom{j_3+|m_1|-|m_2|}{j_3-|m_1|-|m_2|},\; and\; L_{n_2}^{2|m_2|}(0) = \binom{j_2+|m_2|}{j_2-|m_2|} \quad (3.10)$$

We find $\sum_{n_1 n_2, m_1, m_2} \left\langle \Phi_{j_1 m_1}, \Phi_{j_2 m_2} \middle| \Psi_{n j_3 m' m} \right\rangle (x)^{|m_1|} L_{n_1}^{2|m_1|}(x) =$

$$\sqrt{\frac{(2j_3+1)(j_1+j_2-j_3)!(j_3+m)!}{(j_1+j_2+j_3+1)!(j_3+m')!(j_3-m')!(j_3-m)!}} (|m_1|+1)_{(j_3-m)} \times (x)^{(j_3-|m_2|)} L_n^{2j_3+1}(x) \quad (3.11)$$

By applying (3.5) and identifying the two sides of (3.11) we find the analytic expression of matrix elements in term of the hypergeometric function $_3F_2(1)$:

$$\left\langle \Phi_{j_1 m_1}, \Phi_{j_2 m_2} \middle| \Psi_{n j_3 m' m} \right\rangle = (-1)^{n+n_1} (j_1 + j_2 - |m_1| - |m_2|)!(j_1 + j_2 - |m_2| + |m_1|) \times$$

$$\sqrt{\frac{(2j_3+1)(j_3+|m_1|+|m_2|)!(j_3+|m_1|-|m_2|)!}{(j_1+j_2-j_3)!(j_1+j_2+j_3+1)!(j_3-|m_1|-|m_2|)!(j_3-|m_1|+|m_2|)!\prod_{i=1,2}(j_i-|m_i|)!(j_i+|m_i|)!}} \times \quad (3.12)$$

$$_3F_2(-j_1 - j_2 + j_3, -j_1 - j_2 - j_3 - 1, -j_2 + |m_2|; -j_1 - j_2 + |m_1| + |m_2|, -j_1 - j_2 - |m_1| + |m_2|, 1)$$

# 4. Elements of the passage matrix and group theory of Laguerre polynomials

We observe that the basis $|2(2dPH)\rangle$ has several developments, among other $|(4dPH)\rangle$, the development found by Vilenkin [7] using the group theory. We also give the analytical expression of a Clebsh-Gordan coefficient which is particularly useful in comparing these developments.

## 4.1 Group theory of Laguerre polynomials and the passage matrix

To determine the passage matrix elements we develop the basis $|2(2dPH)\rangle$ on the base $|(4dPH)\rangle$ using the expression (3.4) and the expression already established by Vilenkin [7] and the comparisons of these expressions.

If we put $l_1 = (j_1 + j_2 - |m_1| - |m_2|)/2$, $l_2 = (j_1 + j_2 + |m_1| + |m_2|)/2$,

And $j = j_2 - |m_2|$, $m = j_2 + |m_2|$, $x = r^2$.

Using the relations $P^l_{(m,m')}(z) = (i)^{m'-m} d^l_{(m,m')}(z)$, $P^l_{m,m'}(z) = P^l_{-m,-m'}(z)$ we write [7]:

$$\sqrt{\prod_{k=1}^{2} \frac{(j_k - |m_k|)!}{(j_k + |m_k|)!}} (ix\sin\theta)^{|m_2|} (x\cos\theta)^{|m_1|} L^{2|m_2|}_{j_2-|m_2|}(x\sin^2\frac{\theta}{2}) L^{2|m_1|}_{j_1-|m_1|}(x\cos^2\frac{\theta}{2}), = \quad (4.1)$$

$$\sum_{k=0}^{\min(2l_1,2l_2)} (-1)^{k-j} \sqrt{\frac{(2l_1-k)!}{(2l_2-k)!}} P^{l_1}_{(j-l_1,k-l_1)}(\cos\theta) P^{l_2}_{(l_2-m,l_2-k)}(\cos\theta) x^{(|m_1|+|m_2|)} L^{2l_2-2l_1}_{2l_1-k}(x)$$

We have $j - l_1 = (-j_1 + j_2 + |m_1| - |m_2|)/2$, $l_2 - m = (j_1 - j_2 + |m_1| - |m_2|)/2$,

By coupling the product of Wigner's D-matrix then using Regge symmetry

$$P^{l_1}_{jj'}(z) P^{l_2}_{kk'}(z) = \sum_{l=|l_1-l_2|}^{l_1+l_2} \langle l_1 l_2; jk | (l_1 l_2) l(j+k) \rangle \langle l_1 l_2; j'k' | (l_1 l_2) l(j'+k') \rangle P^l_{j+j',j'+k'}(z) \quad (4.2)$$

And $\langle l_1 l_2; (j-l_1),(l_2-m) | (l_1 l_2) l(|m_1|-|m_2|) \rangle = \langle j_1^{|m|} j_2^{|m|}, m_1^{|m|} m_2^{|m|} | (j_1^{|m|} j_2^{|m|}) j_3^{|m|} m_3^{|m|} \rangle$

We get

$$\sum_{l=|l_1-l_2|}^{l_1+l_2} \langle j_1^{|m|} j_2^{|m|}, m_1^{|m|} m_2^{|m|} | (j_1^{|m|} j_2^{|m|}) j_3^{|m|} m_3^{|m|} \rangle d^{j_3^{|m|}}_{|m_1|-|m_2|,|m_1|+|m_2|}(\cos\theta) \times$$

$$\sum_{k=0}^{\min(2l_1,2l_2)} (-1)^{k-j} \sqrt{\frac{(2l_1-k)!}{(2l_2-k)!}} \langle l_1 l_2 m'_1 m'_2 | (l_1 l_2) l m' \rangle x^{(l_2-l_1)} L^{2(l_2-l_1)}_{(2l_1-k)}(x) \quad (4.3)$$

With $\quad m'_1 = k - l_1,\ m'_2 = l_2 - k,\ m' = l_2 - l_1$

Comparing this expression with (3.1) shows that we must develop the basis $\{x^{l_3} L^{(2l_3+1)}_{(l_1+l_2-l_3)}(x)\}$ of the space on the basis $x^{l_2-l_1} L^{2(l_2-l_1)}_{(2l_1-k)}$.

Put $\quad x^{l_3} L^{(2l_3+1)}_{(l_1+l_2-l_3)}(x) = \sum_{k=0}^{\min(2l_1,2l_2)} a_k x^{l_2-l_1} L^{2(l_2-l_1)}_{(2l_1-k)}(x) \quad (4.4)$

We find $\quad a_k = (-1)^{\varphi+k-j} \sqrt{\frac{(2l_1-k)!}{(2l_2-k)!}} \langle l_1 l_2 m'_1 m'_2 | (l_1 l_2) l m' \rangle \quad (4.5)$

Conversely: We start from the expression (3.1) and we use (4.3) and (4.4) we find the expression (4.1).

## 4.2 Expression of the coefficient $\langle l_1, l_2; m'_1, m'_2 | (l_1 l_2) l m' \rangle$

In the expression (4.3) we use the development (3.3) of Laguerre polynomial and then (3.5) we obtain:

$$\sum_{k=0}^{\min(2l_1, 2l_2)} a_k L^{2(l_2-l_1)}_{(2l_1-k)}(x) = \tag{4.6}$$

$$\sum_{\mu=0}^{l_1+l_2-l_3} \sum_{i=0}^{n} (-1)^{\mu+i} \frac{\Gamma(l_1+l_2-l_3+2)(2l_1-\mu)!(2l_2-\mu)!}{\mu!(l_1+l_2-l_3-\mu)!\Gamma(l_3+l_1+l_2-\mu+2)(2l_1-\mu-i)!(2(l_2-l_1)+i)!} L^{2(l_2-l_1)}_i(x)$$

Put $i = 2l_2 - k$ and identify the two sides of (4.6) we find finally the expression of Clebsh-Gordan coefficient.

$$\langle l_1, l_2; k-l_1, l_2+k | (l_1 l_2) l_3, l_2 - l_1 \rangle = (-1)^{\varphi+k-j+2l_2-k}$$

$$\sum_{\mu=0}^{l_1+l_2-l_3} (-1)^{\mu} \frac{\Gamma(l_1+l_2-\mu+1)}{\mu!(l_1+l_2-l_3-\mu)!\Gamma(l_3+l_1+l_2-\mu+2)} \times \frac{(2l_2-\mu)!}{(k-\mu)!} \tag{4.7}$$

By identifying also the expression (4.3) with (3.1) and using the orthogonality of Laguerre polynomials we find the integral representation of the coefficient:

$$\langle l_1 l_2 m'_1 m'_2 | (l_1 l_2) l m' \rangle = (-1)^{k-j+\varphi} \sqrt{\frac{(2l_2-k)!}{(2l_1-k)!}} \int_0^\infty x^{(m+l)} L^{2(m)}_{(j_1+j_2-k)-(m)}(x) L^{2l+1}_n(x) dx \tag{4.8}$$

## 5. Generating functions of passage elements from 2(2dPH) to 4-dimensions polar basis in terms of magnetic moments (general case)

We build first a new generating function of the polar basis $|(2dPH)\rangle$ and the generating functions of $|4dPH\rangle$. We derive from the overlap of these functions the generating functions of matrix elements in term of magnetic moments (general case).

### 5. 1 The new generating function for the polar basis

We know that the Cartesian basis of harmonic oscillator in Dirac notations is

$$|n_x, n_y\rangle = a_x^{+nx} a_y^{+ny} |0,0\rangle \tag{5.1}$$

This ket is not the eigenfunctions of $L_z$. Thus to obtain the basis which has this property we must take the transformation [8]

$$A_1^+ = \frac{\sqrt{2}}{2}(a_x^+ - i a_y^+), \quad A_2^+ = \frac{\sqrt{2}}{2}(a_x^+ + i a_y^+),$$

$$L_z = (N_1 - N_2), \quad N = N_1 + N_2, \tag{5.2}$$

$$N_1 = A_1^+ A_1, \quad N_2 = A_2^+ A_2$$

The new basis $|N_1, N_2\rangle$ can be written in the form

$$|N_1, N_2\rangle = |j+m, j-m\rangle = A_1^{+j+m} A_2^{+j-m} |0,0\rangle. \tag{5.3}$$

This basis is function of $L_z$ and $N$ with the values 2m and 2j.
The new generating function may be written in the form:

$$|F(t_1, t_2)\rangle = \exp[t_1 A_1^+ + t_2 A_2^+]|0,0\rangle$$

$$= \exp[a_x^+ \sqrt{2}(t_1+t_2)/2 + i a_x^+ \sqrt{2}(-t_1+t_2)/2]|0,0\rangle \tag{5.4}$$

In term of Cartesian coordinates we derive the generating function:

$$F(t,\rho,\varphi) = \frac{1}{\sqrt{\pi}}\exp[-\frac{x^2+y^2}{2}+t_1(x+iy)+t_2(x-iy)-t_1t_2] =$$

$$\sum_{jm}(-1)^{j-m}\frac{t_1^{(j+m)}t_2^{(j-m)}}{\sqrt{(j+m)!(j-m)!}}\Phi_{jm}(\rho,\alpha) \tag{5.5}$$

And $\varphi_{jm}(t) = \frac{t_1^{(j+m)}t_2^{(j-m)}}{\sqrt{(j+m)!(j-m)!}}$ is the basis of Fock-Bargmann space [15, 17] with measure $d\mu(t_1,t_2) = d\mu(t_1)d\mu(t_2)$ and $d\mu(t) = \frac{1}{\pi}\exp(-t\bar{t})dx_t dy_t$, $t = x_t + iy_t$.

### 5.2 The generating function of the 4-dimensions polar basis (4dPH)

The generating function of the basis (4dPH) may be deduced from the generating function of Wigner's D-matrix and the generating function of Laguerre polynomials

#### 5.2.1 The generating function of Wigner's D-matrix

The Schwinger generating function [2] of Wigner's D-matrix is:

$$D(x,z,y) = \exp\left((\xi_1\ \xi_2)\begin{pmatrix}\bar{z}_1 & z_2 \\ -\bar{z}_2 & z_1\end{pmatrix}\begin{pmatrix}\eta_1 \\ \eta_2\end{pmatrix}\right) = \sum_{j,m,m'}\varphi_{jm}(\eta_1,\eta_2)\varphi_{jm'}(\xi_1,\xi_2)D^j_{(m',m)}(\Omega)$$

$$z_1 = r\sin\frac{\theta}{2}e^{i(\psi+\varphi)/2}, \quad z_2 = r\sin\frac{\theta}{2}e^{i(\psi-\varphi)/2}, \quad \Omega = (\psi\theta\varphi), \tag{5.6}$$

$$|z_1| = \rho_1, |z_2| = \rho_2$$

#### 5.2.2 Generating function of Laguerre polynomials

The generating function of Laguerre polynomials, [18-19] is:

$$\sum_{n=0}^{\infty} s^n L_n^{(\alpha)}(r) = \frac{1}{(1-s)^{\alpha+1}} e^{-\frac{s}{1-s}r}$$

We deduce that

$$\frac{1}{(1-s)^{2j_3+2}}\exp[-r^2\frac{s}{(1-s)}] = \sum_{n=0}^{\infty} s^n L_n^{2j_3+1}(r^2) \tag{5.7}$$

#### 5.2.3 The generating function of the basis (4dPH)

Using the above generating functions we find

$$G(s,\xi,\eta,\vec{r}) = [\sum_{j_3,m'm}\varphi_{j_3m'}(\xi_1,\xi_2)\varphi_{j_3m}(\eta_1,\eta_2)(\sum_{n=0}^{\infty}s^n r^{2j_3}L_n^{2j_3+1}(r^2))\times$$

$$D^{j_3}_{(m',m)}(\Omega)] = \sum_{j_3,m'm}\sum_{n=0}^{\infty}K_{nj_3m'}(s,\xi)\varphi_{j_3m}(\eta_1,\eta_2)\Psi_{n,j_3m'm}(r,\psi\theta\varphi) \tag{5.8}$$

With $\quad K_{nj_3m'}(s,\xi) = \frac{2\pi}{N_{n,j_3}}s^n\varphi_{j_3m'}(\xi_1,\xi_2)$

So we get the generating function

$$G(s,\xi,\eta,\vec{r}) = \frac{1}{(1-s)^2}\exp[-r^2\frac{(1+s)}{2(1-s)} + \frac{1}{1-s}[\xi_1(\eta_1\bar{z}_1+\eta_2 z_2)+\xi_2(-\eta_1\bar{z}_2+\eta_2 z_1)]] \tag{5.9}$$

### 5.3 Generating function of passage matrix elements.

The generating function of the passage matrix elements $\langle\Psi_{(j_1j_2)j_3m_3}|\Phi_{j_1m_1},\Phi_{j_2m_2}\rangle$ may be written in the form:

$$G(s,\xi,\eta,t,t') = \int G(s,\xi,\eta,\vec{r})F(t,\rho_1,\varphi_1)F(t',\rho_2,\varphi_2)d\vec{r}$$

$$= \sum_{j_1 j_2 m_1 m_2} \sum_{j_3, m'm} \sum_{p=0}^{\infty} \frac{2\pi^2}{\sqrt{(2j_3+1)}} \frac{\sqrt{(2j_3+p+2)!}}{\sqrt{(p)!}} s^p \times \qquad (5.10)$$

$$\varphi_{j_3 m}(\eta_1, \eta_2) \varphi_{j_3 m'}(\xi_1, \xi_2) \phi_{j_1 m_1}(t) \phi_{j_2 m_2}(t') \langle \Psi_{p, j_3 m'm} | \Phi_{j_1 m_1}, \Phi_{j_2 m_2} \rangle$$

A simple calculation gives the generating functions

$$G(s, \xi, \eta, t, t') = \exp[\xi_1(\eta_1 t_2 + \eta_2 t'_1) + \xi_2(-\eta_1 t'_2 + \eta_2 t_1) - s(t_1 t_2 + t'_1 t'_2)] \qquad (5.11)$$

this expression is the generating functions of the 3j symbols of Schwinger [2].
 We find after development of (5.11) and the identification with (5.10) the expression of the elements matrix of passage in terms of magnetic moments (general case).

$$\langle \Psi_{(j_1 j_2) j_3 m_3} | \Phi_{j_1 m_1}, \Phi_{j_2 m_2} \rangle = (-1)^{j_2 - m_2} \sqrt{(2j_3+1)} \langle j_1^m j_2^m, m_1^m m_2^m | (j_1^m j_2^m) j_3^m m_3^m \rangle$$

With
$$j_1^m = (j_1 + j_2 - m_1 + m_2)/2, \quad m_1^m = (j_2 - j_1 + m_1 - m_2)/2,$$
$$j_2^m = (j_1 + j_2 + m_1 - m_2)/2, \quad m_2^m = (-j_2 + j_1 + m_1 + m_2)/2,$$
$$j_3^m = j_3, \quad m_3^m = (m_1 + m_2) \qquad (5.12)$$

We find also the interesting transformation
$$G(s, \xi, \eta, \vec{r}) = \int G(s, \xi, \eta, t, t') \overline{F(t, \rho_1, \varphi_1)} \, \overline{F(t', \rho_2, \varphi_2)} d\mu(t) d\mu(t') \qquad (5.13)$$

## 6. The symmetry of 3j symbols and the new generating functions of Clebsh-Gordan coefficients

We give first the symmetry coefficients Clebsh-Gordan, result of the preceding paragraphs, and then the new generating function of these coefficients which is a function of two parameters only.

**6.1 Symmetry of 3j symbols**

Comparing the expressions (3.7) and (5.12) we obtain two equivalent expressions of Clebsh-Gordan coefficients:

$$\langle j_1^m j_2^m, m_1^m m_2^m | (j_1^m j_2^m) j_3^m m_3^m \rangle = \langle j_1^{|m|} j_2^{|m|}, m_1^{|m|} m_2^{|m|} | (j_1^{|m|} j_2^{|m|}) j_3^{|m|} m_3^{|m|} \rangle$$

We find four expressions of symmetry [16-17]

$$|m_1| = m_1, \; |m_2| = m_2, \qquad |m_1| = -m_1, \; |m_2| = m_2,$$
$$|m_1| = m_1, \; |m_2| = -m_2, \qquad |m_1| = -m_1, \; |m_2| = -m_2. \qquad (6.1)$$

We made a computer program (Maple) to make the verification.
Put $\quad J_2 - J_1 = m_1 - m_2, \quad J_1 + J_2 = j_1 + j_2$
With $\quad m_1 = (M - J_1 + J_2)/2, \; m_2 = (M + J_1 - J_2)/2, \; M = m_1 + m_2 = M_1 + M_2$
We obtain also in terms of 3j symbols the expression:

$$\begin{pmatrix} J_1 & J_2 & j_3 \\ M_1 & M_2 & -M \end{pmatrix} = \begin{pmatrix} (J_1 + J_2 - |m_1| + |m_2|)/2 & (J_1 + J_2 + |m_1| - |m_2|)/2 & j_3 \\ (M_1 - M_2 + |m_1| + |m_2|)/2 & (-M_1 + M_2 + |m_1| + |m_2|)/2 & -M \end{pmatrix}$$

**6.2 The new generating function of Clebsh-Gordan coefficients**

Multiplying the two sides of the expression (3.11) by $(x)^{|m_1|+|m_2|} L_{n_1}^{2|m_1|}(x)$ and after integration we find:

$$\langle \Psi_{(j_1 j_2) j_3 m_3} | \Phi_{j_1 m_1}, \Phi_{j_2 m_2} \rangle = A(j,|m|) \int dx (x)^{j_3+|m_1|-|m_2|} L_{n_1}^{2|m_1|}(x) L_n^{2j_3+1}(x)$$

And $A(j,|m|) =$

$$\left[\frac{(2j_3+1)(j_3+|m_1|+|m_2|)!(j_3+|m_1|-|m_2|)!}{(J-2j_3)!(J+1)!(j_3-|m_1|-|m_2|)!(j_3-|m_1|+|m_2|)!} \prod_{i=1,2} \frac{(j_i-|m_i|)!}{(j_i+|m_i|)!}\right]^{1/2} \frac{(2|m_2|)!}{(2|m_1|)!}$$

With the help of the generating function of Laguerre polynomial we derive the new generating function of Clebsh-Gordan coefficients

$$\sum \frac{1}{A(j,|m|)} u^{n_1} v^n \langle \Psi_{(j_1 j_2) j_3 m_3} | \Phi_{j_1 m_1} \Phi_{j_2 m_2} \rangle =$$

$$\int_0^\infty (x)^{(j_3+|m_1|-|m_2|)} \frac{\exp[-xu/(1-u)]}{(1-u)^{2j_3+2}} \frac{\exp[-xv/(1-v)]}{(1-v)^{2|m_1|+1}} e^{-x} dx =$$

$$(j_3+|m_1|-|m_2|)! \frac{(1-u)^{(-j_3+|m_1|-|m_2|-1)}(1-v)^{(j_3-|m_1|+|m_2|)}}{(1-uv)^{(j_3+|m_1|-|m_2|)+1}} \quad (6.2)$$

So we obtain a generating function with two parameters and not three as in the Schwinger's work [2] and the development of this expression gives (3.12).

### 7. Schwinger approach for 6j, 9j symbols and the polar basis of harmonic oscillator

In this part we determine the function of the coupling of two angular momentums in terms of creation operators of the polar basis of oscillator and we demonstrate that the choice of the polar basis gives also the recoupling coefficients 3nj.

**7.1 The generating function of the coupled basis in terms of creation operators**

In this part we choose the polar basis by

$$|jm\rangle = (-1)^{j-m} A_1^{+j+m} A_2^{+j-m} |0,0\rangle \quad (7.1)$$

And the generating function of the elements of passage $\langle 2(2pdPH)|(4dPH)\rangle$ is

$$G(\alpha,\eta,t,t') = \exp[\alpha_1(\eta_1 t_2 + \eta_2 t'_1) + \alpha_2(\eta_1 t'_2 + \eta_2 t_1) + \alpha_3(t'_1 t'_2 - t_1 t_2)]$$

Using Fock-Bargmann space and Schwinger notations we write

$$G(\alpha,x,\bar{a},\bar{b}) = \exp[\alpha(x_1 \bar{a}_1 + x_2 \bar{a}_2) + \beta(x_1 \bar{b}_1 + x_2 \bar{b}_2) + \gamma(\bar{a}_1 \bar{b}_2 - \bar{a}_2 \bar{b}_1)] \quad (7.2)$$

The generating function of the 4-dimension polar basis which is the coupling basis

$$G(\alpha,x,q) = \int G(\alpha,x,\bar{a},\bar{b}) F_{(j_1 m_1)}(a_1 b_2, \rho_1, \varphi_1) F_{(j_2 m_2)}(a_2 b_1, \rho_2, \varphi_2) \prod_{i=1}^{2} d\mu(a,b) \quad (7.3)$$

With $\quad q = (u_1, u_2, u_3, u_4), \alpha = (\alpha, \beta, \gamma)$

Writing this expression in terms of creations and destructions operators:

$$|G(\alpha,x,A)\rangle = \int \{G(\alpha,x,\bar{a},\bar{b}) | F^{12}(a,b)\rangle\} d\mu(a) d\mu(b) \quad (7.4)$$

$F^{12}(a,b)$ Is the generating function of the uncoupled basis.

With $\quad |F^{12}(a,b)\rangle = \exp(b_2 A_1^+ - a_1 A_4^+ + a_2 A_3^+ - b_1 A_2^+)|0\rangle$

We find also

$$|G(\alpha,\eta,A)\rangle = \exp[\xi_1(-\eta_1 A_4^+ + \eta_2 A_3^+) + \xi_2(-\eta_1 A_2^+ + \eta_2 A_1^+) + s(A_1^+ A_4^+ - A_2^+ A_3^+)]|0\rangle$$

And $\quad G(\alpha,\eta,\vec{r}) = \langle r\psi\theta\varphi | G(\alpha,\eta,A)\rangle \quad (7.5)$

The development of this expression gives the expression of the function (2.10) in terms of the creation and destruction operators of harmonic oscillator. This result is similar to the well known three-dimensional case of harmonic oscillator [22].

**7.2 The coupling of three and four angular momentum**

To determine the expressions of Racah coefficients and 9j symbols we follow Schwinger's method for the coupling of angular momentum.

We write the coupling of four angular momentums by two different ways.

$$\vec{J} = (\vec{J}_1 + \vec{J}_2) + (\vec{J}_3 + \vec{J}_4) = (\vec{J}_1 + \vec{J}_4) + (\vec{J}_2 + \vec{J}_3) \tag{7.6}$$

Put $\quad J_1 = (j_1 j_2)j_{12}, J_2 = (j_3 j_4)j_{34}, J'_1 = (j_1 j_4)j_{14}, J'_2 = (j_2 j_3)j_{24}$

And to be the clearest possible we write

$$G(\alpha, x(ab), q) = \int G(\alpha, x, \bar{a}, \bar{b}) F_{(j_1 m_1)}(a_1 b_2, \rho_1, \varphi_1) F_{(j_2 m_2)}(a_2 b_1, \rho_2, \varphi_2) d\mu(a,b) \tag{7.7}$$

We denote the generating function of the first coupling by

$$G_{J_1, J_2}(\alpha, z; q_1, q_2) = \int G(\alpha_3, z, \bar{x}, \bar{y}) G_{12}(\alpha_1, x(ab), q_1) G_{34}(\alpha_2, y(cd), q_2) d\mu(x,y)) \tag{7.8}$$

$$= \int G_{J_1, J_2}(\alpha, z, abcd) \{ F_{(j_1 m_1)}(a_1 b_2, \rho_1, \varphi_1) F_{(j_2 m_2)}(a_2 b_1, \rho_2, \varphi_2)$$

$$\times F_{(j_3 m_3)}(c_1 d_2, \rho_3, \varphi_3) F_{(j_4 m_4)}(c_2 d_1, \rho_4, \varphi_4) \} d\mu(a,b,c,d)$$

With $\quad G_{J_1, J_2}(\alpha, z, abcd) = e^{Q(\alpha, z, abcd)}$,

$$Q(\alpha, z, abcd) = \alpha_3 [ab] + \beta_3 [cd] + \gamma_3 \alpha_1 \beta_1 [[bd] + \gamma_3 \alpha_1 \beta_2 [bc] + \gamma_3 \alpha_2 \beta_1 [ad]$$
$$+ \gamma_3 \alpha_2 \beta_2 [ac] + \gamma_2 \alpha_2 (za) + \gamma_2 \alpha_1 (zb) + \gamma_1 \beta_2 (zc) + \gamma_1 \beta_1 (zd) \tag{7.9}$$

And $\quad [ab] = a_1 b_2 - a_2 b_1, (ab) = a_1 b_1 + a_2 b_2$

$$q_1 = (u_1, u_2, u_3, u_4), q_2 = (u_5, u_6, u_7, u_8), \alpha_i = (\alpha_i, \beta_i, \gamma_i)$$

The generating function of the second coupling is

$$G_{J'_1, J'_2}(\alpha', z; q'_1, q'_2) = \int G_{J'_1, J'_2}(\alpha', z, adcb) \{ F_{(j_1 m_1)}(a_1 d_2, \rho_1, \varphi_1) F_{(j_2 m_2)}(a_2 d_1, \rho_2, \varphi_2)$$

$$\times F_{(j_3 m_3)}(c_1 b_2, \rho_3, \varphi_3) F_{(j_4 m_4)}(c_2 b_1, \rho_4, \varphi_4) \} d\mu(a', b', c', d')$$

And $\quad q'_1 = (u_1, u_2, u_7, u_8), q'_2 = (u_5, u_6, u_3, u_4), \alpha'_i = (\alpha'_i, \beta'_i, \gamma'_i)$

The generating function of the recoupling coefficient $\langle (J_1 J_2) j | (J'_1 J'_2) j \rangle$ is:

$$G(\alpha, \alpha'; z, \bar{z}) = \int [G_{J'_1, J'_2}(\alpha', \bar{z}; q'_1, q'_2) G_{J_1, J_2}(\alpha, z; q_1, q_2) \prod_{i=1}^{8} du_i \tag{7.10}$$

But $\quad \int \langle F^{14}(a', d') F^{32}(c', b') | F^{12}(a,b) F^{34}(c,d) \rangle d\mu(a',b',c',d') \} = e^{\bar{a}a' + \bar{b}b' + \bar{c}c' + \bar{d}d'}$

In Fock-Bargmann space we use the formula

$$f(\bar{z}) = \int e^{\bar{z}z'} f(\bar{z}') d\mu(z') \tag{7.11}$$

We find finally

$$G(\langle (J_1 J_2) j | (J'_1 J'_2) j \rangle) = \int [G_{J'_1, J'_2}(\alpha', \bar{z}, \bar{a}\bar{d}\bar{c}\bar{b}) G_{J_1, J_2}(\alpha, z, abcd) d\mu(a,b,c,d,z) \tag{7.12}$$

The calculation of this expression is done by means of Bargmann integral [15, 17] and it is clear that we get the same Schwinger generating functions of 6j and 9j symbols. The generalization of this result to recoupling coefficients 3nj may be done by the same method.

**Acknowledgments**

Because I have been retired since 2006 I have to thank everyone who helped me through this life. I also thank Professor M. Kibler from the IPN-Lyon for his encouragements.